\documentclass[superscriptaddress,preprintnumbers,showpacs,prb,aps,twocolumn]{revtex4}

\usepackage{amsfonts}
\usepackage{amssymb}
\usepackage{amsmath}
\usepackage{graphicx}
\usepackage{natbib}
\usepackage{physics}
\usepackage{siunitx}
\usepackage{float}
\usepackage{multirow}

\usepackage{fancyhdr}

\setcitestyle{numbers,square}

\usepackage[usenames,dvipsnames]{color}
\usepackage{soul}

\usepackage[hidelinks]{hyperref} 

\usepackage{tablefootnote}
\makeatletter
\newcommand\footnoteref[1]{\protected@xdef\@thefnmark{\ref{#1}}\@footnotemark}
\makeatother

\begin{document}

\newcommand{\ie}{{\it i.e.}}
\newcommand{\eg}{{\it e.g.}}
\newcommand{\etal}{{\it et al.}}

\newcommand{\micron}{$\mu$m}

\newcommand{\Kxx}{$\kappa_{\rm {xx}}$}
\newcommand{\Kxy}{$\kappa_{\rm {xy}}$}
\newcommand{\Kzy}{$\kappa_{\rm {zy}}$}
\newcommand{\Kzz}{$\kappa_{\rm {zz}}$}

\newcommand{\ncco}{Nd$_{2-x}$Ce$_x$CuO$_4$}
\newcommand{\pcco}{Pr$_{2-x}$Ce$_x$CuO$_4$}

\newcommand{\TN}{$T_{\rm {N}}$}
\newcommand{\Tc}{$T_{\rm {c}}$}

\title{Thermal Hall conductivity of electron-doped cuprates}

\author{Marie-Eve~Boulanger}
\affiliation{Institut quantique, D\'epartement de physique \& RQMP, Universit\'e de Sherbrooke, Sherbrooke, Qu\'ebec, J1K 2R1, Canada}

\author{Ga\"el~Grissonnanche}
\affiliation{Institut quantique, D\'epartement de physique \& RQMP, Universit\'e de Sherbrooke, Sherbrooke, Qu\'ebec, J1K 2R1, Canada}

\author{\'Etienne~Lefran\c cois}
\affiliation{Institut quantique, D\'epartement de physique \& RQMP, Universit\'e de Sherbrooke, Sherbrooke, Qu\'ebec, J1K 2R1, Canada}

\author{Adrien~Gourgout}
\affiliation{Institut quantique, D\'epartement de physique \& RQMP, Universit\'e de Sherbrooke, Sherbrooke, Qu\'ebec, J1K 2R1, Canada}

\author{Ke-Jun~Xu}
\affiliation{Geballe Laboratory for Advanced Materials, Stanford University, Stanford, California, 94305, USA}
\affiliation{Stanford Institute for Materials and Energy Sciences, SLAC National Accelerator Laboratory, Menlo Park, California, 94025, USA}
\affiliation{Departments of Physics and Applied Physics, Stanford University, Stanford, California, 94305, USA}

\author{Zhi-Xun~Shen}
\affiliation{Geballe Laboratory for Advanced Materials, Stanford University, Stanford, California, 94305, USA}
\affiliation{Stanford Institute for Materials and Energy Sciences, SLAC National Accelerator Laboratory, Menlo Park, California, 94025, USA}
\affiliation{Departments of Physics and Applied Physics, Stanford University, Stanford, California, 94305, USA}

\author{Richard~L.~Greene}
\affiliation{Department of Physics, Maryland Quantum Materials Center, University of Maryland, College Park, Maryland 20742, USA}

\author{Louis~Taillefer}
\affiliation{Institut quantique, D\'epartement de physique \& RQMP, Universit\'e de Sherbrooke, Sherbrooke, Qu\'ebec, J1K 2R1, Canada}
\affiliation{Canadian Institute for Advanced Research, Toronto, Ontario M5G 1M1, Canada}

\date{\today}

\begin{abstract}
Measurements of the thermal Hall conductivity in hole-doped cuprates 
have shown that phonons acquire chirality in a magnetic field, both in the pseudogap phase and in the Mott insulator state. 
The microscopic mechanism at play is still unclear.
A number of theoretical proposals are being considered, including skew scattering of phonons by various defects, 
the coupling of phonons to spins, and a state of loop-current order with the appropriate symmetries, 
but more experimental information is required to constrain theoretical scenarios. 
Here we present our study of the thermal Hall conductivity \Kxy~in the electron-doped cuprates \ncco~and \pcco, 
for dopings across the phase diagram, from $x$ = 0, in the insulating antiferromagnetic phase, up to $x$ = 0.17, in the metallic phase above optimal doping. 
We observe a large negative thermal Hall conductivity at all dopings, in both materials.
Since heat conduction perpendicular to the CuO$_2$ planes is dominated by phonons, the large thermal Hall conductivity we observe in electron-doped cuprates for a heat current in that direction must also be due to phonons, as in hole-doped cuprates.
%
However, the degree of chirality, measured as the ratio $|$\Kxy/\Kxx$|$, where \Kxx~is the longitudinal thermal conductivity,
is much larger in the electron-doped cuprates.
We discuss various factors that may be involved in the mechanism that confers chirality to phonons in cuprates, including short-range spin correlations.
\end{abstract}

\pacs{Valid PACS appear here}
\maketitle

\section{INTRODUCTION}

The thermal Hall effect is a promising
tool for obtaining information on the nature of excitations in quantum materials. 
The thermal Hall conductivity \Kxy~is obtained by measuring the transverse temperature difference (along the $y$ axis) 
produced by a magnetic field applied along the $z$ axis, 
perpendicular to the heat current (along the $x$ axis).
Electrons give rise to a thermal Hall signal in metals, due to the Lorentz force,
for the same reason they produce an electrical Hall effect.
In recent years, it has become clear that insulators can also give rise to a thermal Hall signal,
even though their excitations are charge neutral.
Magnons can produce a thermal Hall effect under certain conditions~\cite{Lee2015},
as observed in the antiferromagnet Lu$_2$V$_2$O$_7$~\cite{Onose2010}.
It has been proposed that more exotic excitations, such as Majorana edge modes in Kitaev spin liquids~\cite{Kasahara2018}, 
can also be detected via the thermal Hall effect.

Phonons can also give rise to a thermal Hall signal, even if they are {charge} neutral excitations.
The first observation of a phonon thermal Hall effect was in the paramagnetic 
insulator Tb$_3$Ga$_5$O$_{12}$~\cite{Strohm2005,Inyushkin2007}. 
In that case, the effect was attributed to the skew scattering of phonons by superstoichiometric Tb$^{3+}$ ions~\cite{Mori2014}. 
Since then, a phonon Hall effect has been reported in various other insulators.
In the multiferroic material Fe$_2$Mo$_3$O$_8$, 
the effect was linked to the strong spin-lattice interaction~\cite{Ideue2017}.
In the quantum paraelectric SrTiO$_3$, 
the effect was linked to a strong flexoelectric susceptibility and the presence of structural domains~\cite{Li2020,Chen2020}.
In the antiferromagnet Cu$_3$TeO$_6$,
the record-high \Kxy~signal was linked to a large phonon conductivity \Kxx, giving a ratio $|$\Kxy/\Kxx$|$ comparable to that measured in other insulators~\cite{Chen2021}.

In hole-doped cuprates, a large negative \Kxy~signal was observed in the pseudogap phase below the critical doping $p^\star$~\cite{Grissonnanche2019},
but also in the Mott insulator state at zero doping, 
in La$_2$CuO$_4$, Nd$_2$CuO$_4$ and Sr$_2$CuO$_2$Cl$_2$~\cite{Boulanger2020}.

In La$_2$CuO$_4$, 
phonons were shown to be responsible for the thermal Hall effect
by applying a heat current normal to the CuO$_2$ planes, 
a direction in which only phonons can move easily, 
which revealed that the $c$-axis thermal Hall conductivity, \Kzy, 
is comparable in magnitude to the in-plane conductivity, \Kxy~\cite{Grissonnanche2020}.

The question is this: by what mechanism can a magnetic field confer chirality -- here defined as handedness in a magnetic field -- 
to phonons, and thus produce a thermal Hall effect,
in cuprates, and more generally in insulators?
A number of theoretical proposals have recently been made, 
including the scattering of phonons by defects~\cite{Guo2021,flebus2021,sun2021,guo2022}
, 
the coupling of phonons to spins~\cite{ye2021} 
and a state of loop-current order with the appropriate symmetries~\cite{Varma2020}. 
It is currently unclear what precise mechanism is appropriate for the cuprates, or indeed for Fe$_2$Mo$_3$O$_8$, SrTiO$_3$ and Cu$_3$TeO$_6$. 
%
%

\begin{figure}[t]
\centering
\includegraphics[width = \linewidth]{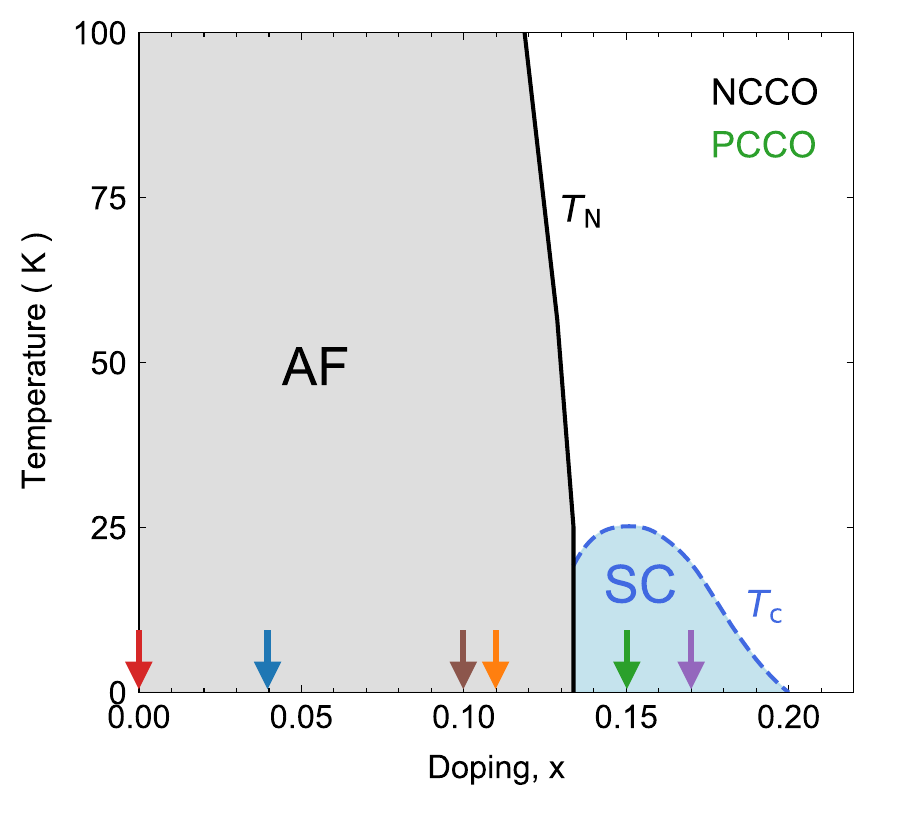}
\caption{
Schematic temperature-doping phase diagram of the electron-doped cuprates, 
consisting of the antiferromagnetic phase (AF), bounded by the N\'eel temperature \TN~(solid black line),
and the superconducting phase (SC),
bounded by its zero-field critical temperature \Tc~(dashed blue line).
In our study, 
we applied a magnetic field of $H$ = 15~T normal to the CuO$_2$ planes
in order to remove the superconductivity and access the normal state down to $T \to 0$.
The color-coded arrows indicate the doping values of the samples included in our study:
Nd$_2$CuO$_4$ ($x=0$),
NCCO ($x$ = 0.04, 0.10, 0.11, 0.17)
and 
PCCO ($x$ = 0.15; green arrow). We divide the doping range into an AF regime ($x<0.13$) and a metallic regime ($x>0.13$).
}
\label{fig1}
\end{figure}

To shed new light on the thermal Hall effect in cuprates, 
we have conducted a systematic study of the thermal Hall conductivity in the electron-doped cuprates \ncco~(NCCO) and \pcco~(PCCO). 
There are two important differences between electron-doped and hole-doped cuprates:
the mysterious pseudogap phase of the latter,
{characterized by a partial spectral weight depletion in the antinodal region of the Fermi surface~\cite{Matt2015},} is not present in the former~\cite{Armitage2010},
and antiferromagnetism (AF) is stronger in the former, in the sense that the phase of long-range AF order
extends to much higher doping~\cite{motoyama2007} -- up to $x \simeq 0.13$ in NCCO (Fig.~\ref{fig1}) vs $p \simeq 0.02$ in La$_{2-x}$Sr$_x$CuO$_4$ (LSCO).
In addition, electron-doped cuprates are convenient because the full doping range can be covered,
from Mott insulator at zero doping to metal at high doping, while keeping the same crystal structure,
and a modest magnetic field ($\simeq 10$~T) is sufficient to fully suppress superconductivity~\cite{Tafti2014},
so that the normal state can easily be accessed down to $T \to 0$ at all dopings. 
%
%
%

%
%

In this paper, 
we show that there is a large negative thermal Hall conductivity 
in the electron-doped cuprates that persists up to the highest doping we have investigated ($x = 0.17$).
Because a large thermal Hall signal is still observed when the heat current is applied normal to the CuO$_2$ planes,
we conclude that phonons are responsible for it. 
Indeed, all other heat-carring excitations (of electronic or magnetic origin) are expected to be much more mobile within the CuO$_2$ planes than across planes~\cite{Grissonnanche2020}.
The fact that $|$\Kxy$|$~remains large outside the phase of long-range AF order ($p >0.13$, Fig.~\ref{fig1}) 
shows that AF order {\it per se} is not an important ingredient in the mechanism that confers chirality to phonons in a magnetic field.
The fact that the signal is very similar for NCCO and PCCO rules out rare-earth ions as playing a key role.
The phononic \Kxy~response in electron-doped cuprates has the same sign (negative) and a similar temperature dependence 
to that found in hole-doped cuprates, but with a significantly larger magnitude.
Indeed, the degree of chirality, 
measured as the ratio $|$\Kxy/\Kxx$|$, where \Kxx~is the longitudinal thermal conductivity,
is $\sim 4$ times larger in the electron-doped cuprates.
We infer that the same underlying mechanism is effective in all cuprates 
and suggest that short-range AF correlations may play a role.

\section{METHODS}

\subsection{Samples}

Single crystals of \ncco~with $x = 0.04$, 0.10, 0.11 and 0.17 were grown at Stanford University by the traveling-solvent floating-zone method in O$_2$ and annealed in flowing argon for 48 hours at 900$^{\circ}$C.
A single crystal of \pcco~with $x=0.15$ was grown at the University of Maryland using the flux growth method with
flowing argon and a titanium getter at 900$^{\circ}$ for 4 days~\cite{Peng1991}.
The value of $T_c$, the superconducting transition temperature in zero field, defined by the onset of the drop in magnetization, 
is $T_c = 22$~K and 20~K for $x$ = 0.15 and 0.17, respectively.
For these superconducting samples, a field of 15 T applied normal to the CuO$_2$ planes is large enough 
to completely suppress superconductivity down to $T \to 0$~\cite{Tafti2014}.

Samples were cut into rectangular platelets and contacts were made 
with silver epoxy, diffused at 500\si{\celsius} under oxygen for 1~\si{\hour}.
The in-plane samples 
have their lengths (and current direction) along the $a$ axis ($J \parallel a \parallel x$), 
\ie~parallel to the CuO$_2$ planes of the tetragonal structure. 
%
%
Sample dimensions are roughly $(500-1000) \times (500-1000) \times (50-200)$ (length between contacts $\times $ width $\times$ thickness, in $\mu$m).

We also studied thermal transport 
for heat current applied along the $c-$axis ($J \parallel c \parallel z$), for 
NCCO at $x = 0.04$ and $x=0.17$. 
To do so, 
we cut two samples from the same single crystal (distinct from the crystal used for the in-plane transport study):
one with the longest direction along the $c$ axis and the other along the $a$ axis. 
%
%
%

In addition to the annealed samples mentioned above,
we also measured two as-grown (unannealed) samples, with $x$ = 0.04 and 0.10, respectively -- 
to investigate the effect of oxygen reduction on the thermal transport properties of NCCO.
The dimensions are ($1320 \times 1250 \times 150$) $\mu$m and ($700 \times 550 \times 670$) $\mu$m for $x = 0.04$ and $x = 0.10$ respectively.
For $x <0.13$,
the reduction process 
removes some of the non-stoichiometric excess oxygen atoms that lie naturally in apical positions of the structure in as-grown samples~\cite{Richard2004}.
The effectiveness of this reduction process decreases with increasing Ce doping~\cite{Armitage2002}. 
%
%

\subsection{Measurements}
The thermal conductivity \Kxx~is measured by applying a heat current $J_{x}$ along the $x$ axis of the sample (longest direction), 
which generates a longitudinal temperature difference $\Delta T_{\rm{x}} = T^{+} - T^{-}$. 
The thermal conductivity \Kxx~is given by
\begin{align}
  \kappa_{\rm{xx}} = \frac{J_{x}}{\Delta T_{\rm x}}\left(\frac{L}{wt}\right),
\end{align}
{where $J_x$ is the heat current,} $w$ is the sample width, $t$ its thickness and $L$ the distance between $T^{+}$ and $T^{-}$.
%
When a magnetic field is applied parallel to $z$, 
a transverse temperature difference $\Delta T_y$ can develop along the $y$ axis. 
The thermal Hall conductivity \Kxy~is then defined as 
\begin{align}
  \kappa_{\rm{xy}} = -\kappa_{\rm{yy}}\left(\frac{\Delta T_{y}}{\Delta T_{x}}\right)\left(\frac{L}{w}\right).
\end{align}
In a tetragonal system (like NCCO and PCCO), 
we can take $\kappa_{\rm{yy}} = \kappa_{\rm{xx}}$. 
The temperature differences $\Delta T_{\rm{x}}$ and $\Delta T_{\rm{y}}$ were measured using
type-E thermocouples (chromel-constantan), 
in a steady-state method at fixed magnetic field $H$ (see refs.~\onlinecite{Boulanger2020,Grissonnanche2020,Grissonnanche2016}),
since such thermocouples have a weak field dependence in the regime of $T$ and $H$ values explored here,
and they have a better sensitivity than resistive Cernox sensors at high temperature.
(Note that in a previous test we found no difference between a \Kxy~measurement using 
thermocouples and a \Kxy~measurement on the same sample using Cernox sensors 
-- see Appendix C in ref.~\onlinecite{Grissonnanche2016}.)
%
%
The (heat-off) voltage background of the thermocouples was carefully subtracted 
from the heat-on signal 
to give the correct $\Delta T_y$.
In order to properly measure \Kxy, 
any contamination from the thermal conductivity \Kxx~that would result from a slight misalignment of the two opposite 
transverse contacts is removed by field anti-symmetrization, namely 
$\Delta T_{\rm y}(H) =\left[\Delta T_{\rm y}(T, H)-\Delta T_{\rm y}(T,-H)\right]/2$.
The heat current along the $x$ axis is generated by a strain gauge heater {attached} at one end of the sample.
The other end is glued using silver paint to a copper block that serves as a heat sink.
(Note that the Hall response of copper in a field does not contaminate the Hall response coming from the sample, 
as was verified by measuring \Kxy~twice on the same sample of Nd$_2$CuO$_4$ -- once 
with the copper block and then with a block made of the insulator LiF.
%
The same \Kxy~curve was obtained with the two setups, within error bars -- see Supplementary information in ref.~\onlinecite{Boulanger2020}.
Hence, using copper for the heat sink does not lead to any detectable contamination of the thermal Hall signal.)
%
%
%
\begin{figure}[t]
\centering{}
\includegraphics[width = 0.8\linewidth]{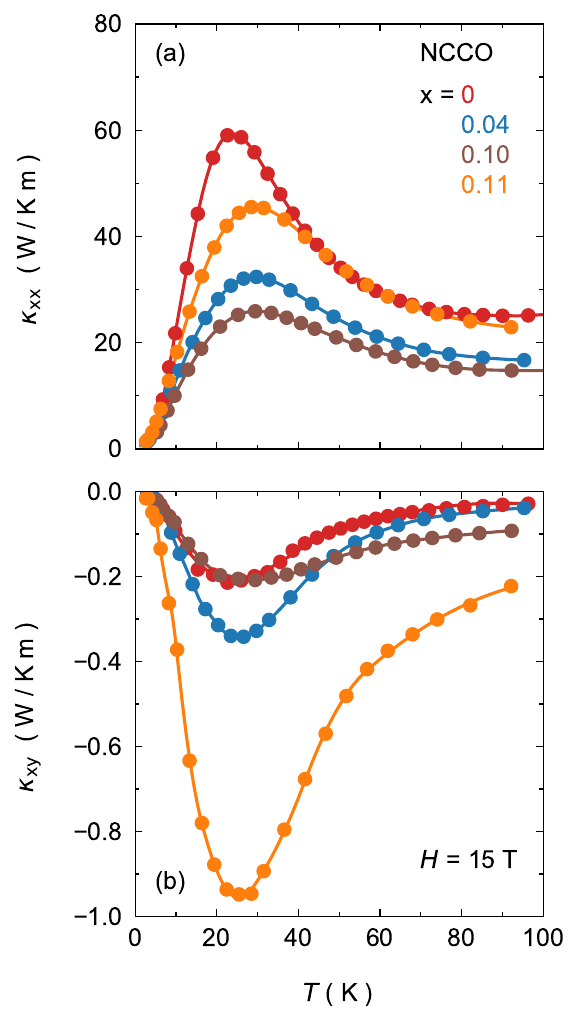}
\caption{
Thermal transport properties of NCCO in the AF regime, at dopings
$x = 0$ (red; \cite{Boulanger2020}), 
0.04 (blue), 
0.10 (brown)
and 0.11 (orange),
measured in a magnetic field $H$ = 15~T applied normal to the CuO$_2$ planes ($H \parallel c$).
(a)~Thermal conductivity \Kxx~vs temperature.
(b)~Thermal Hall conductivity \Kxy~vs $T$. 
{All lines through the data points are a guide to the eye.}
}
\label{fig2}
\end{figure}
%
%

For the measurements of \Kzy~presented here,
the heat current $J$ was sent along the $c$ axis of the single crystal (along $z$), 
perpendicular to the CuO$_2$ planes. 
By applying a magnetic field $H$ along the $a$ axis (along $x$), 
parallel to the CuO$_2$ planes, 
a transverse temperature difference is generated along $y$. 
The longitudinal thermal conductivity along the $c$ axis is then given by 
$\kappa_{\rm{zz}} = \left( J_z / \Delta T_z\right) \left( L / w t\right)$. 
The out-of-plane thermal Hall conductivity is defined as 
$\kappa_{\rm{zy}} = - \kappa_{\rm{yy}}\left( \Delta T_y / \Delta T_z\right) \left( L / w \right)$, 
where $\kappa_{\rm{yy}}$ is the longitudinal thermal conductivity along the $y$ axis,
again taken to be equal to $\kappa_{xx}$ for a tetragonal system. 
More details can be found in ref.~\onlinecite{Grissonnanche2020}.

\section{RESULTS}
%
%
%
%
%
\begin{figure}[t]
\centering
\includegraphics[width = 0.8\linewidth]{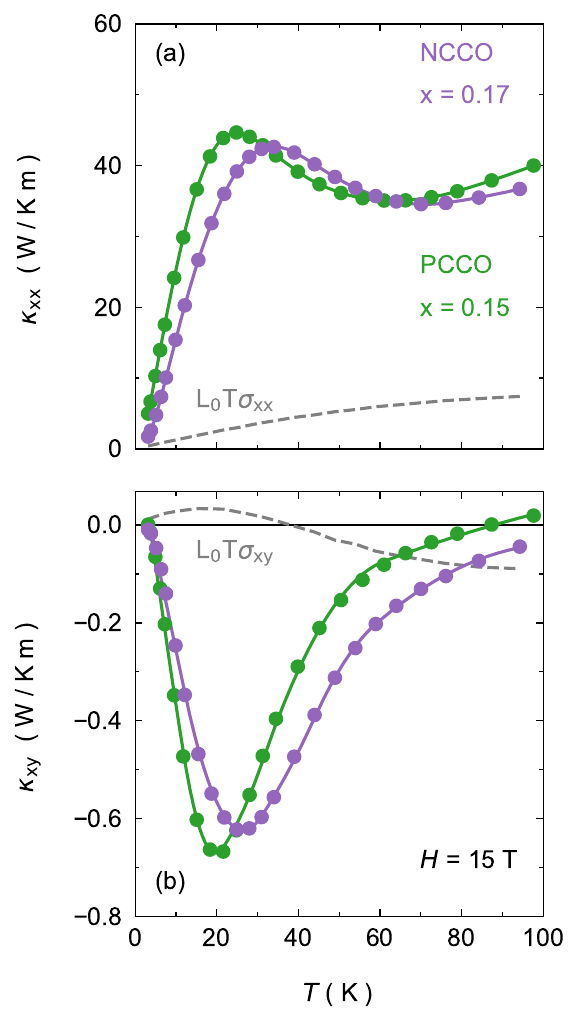}

\caption{
Thermal transport properties of electron-doped cuprates in the {metallic} regime, namely
PCCO at $x = 0.15$ (green) and NCCO at $x = 0.17$ (purple),
measured in a field $H = 15$~T ($H \parallel c$).
(a)~\Kxx~vs $T$.
(b)~\Kxy~vs $T$. 
The dashed grey lines are an estimate of the electronic contribution to \Kxx~and \Kxy, respectively,
based on applying the Wiedemann-Franz law to the electrical conductivities $\sigma_{\rm xx}$ and 
$\sigma_{\rm xy}$ reported for PCCO at $x$ = 0.17~\cite{Tafti2014,Charpentier2010} (see text). 
All solid lines are a guide to the eye.
}
\label{fig3}
\end{figure}
%

The longitudinal thermal conductivity \Kxx~and the thermal Hall conductivity \Kxy~of our samples
are displayed in Fig.~\ref{fig2} (for the AF regime) and Fig.~\ref{fig3} (for the {metallic} regime).
Our main finding is that a large negative thermal Hall signal is present in all samples, at all dopings.
The minimum in \Kxy~is located at roughly the same temperature as the maximum in the phonon-dominated \Kxx,
namely $T \simeq 20-30$~K,
a first indication that the thermal Hall effect is due to phonons.

\begin{figure}[t]
\centering
\includegraphics[width = 0.795\linewidth]{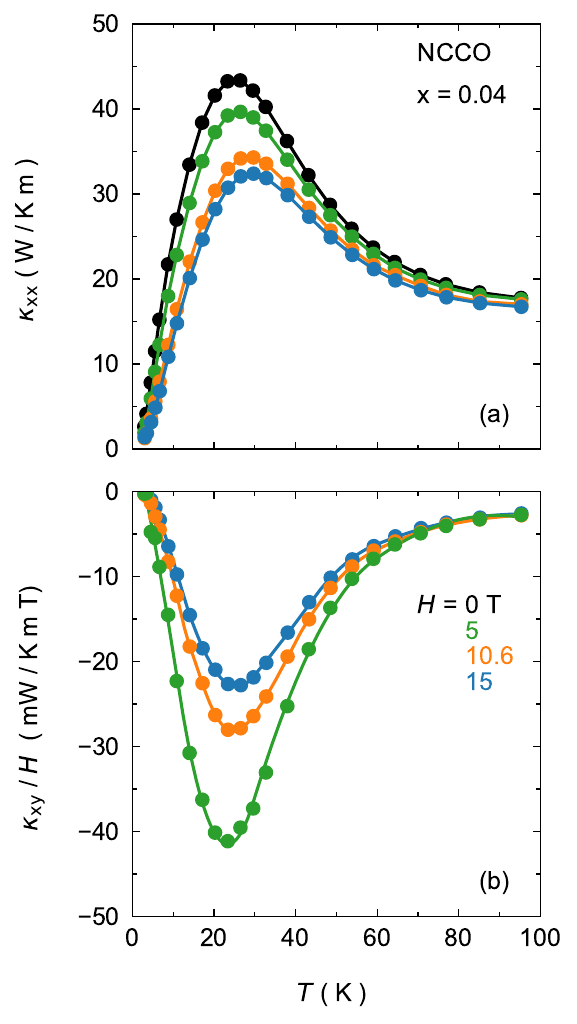}
\caption{
Field dependence of thermal transport in NCCO at $x = 0.04$.
%
(a) Thermal conductivity \Kxx~vs $T$ for $H = 0$ (black), 5~T (green), 10.6~T (orange) and 15~T (blue). 
Near its peak value, \Kxx~is seen to drop significantly with increasing field. 
A similar field-induced suppression of \Kxx~is observed at all dopings.
(b) Thermal Hall conductivity for $H = 5$~T (green), 10.6~T (orange) and 15~T (blue), plotted as \Kxy$/H$ vs $T$. 
Although \Kxy~is linear in $H$ at the highest temperatures
({\it i.e.} \Kxy/$H$ is independent of $H$), 
it acquires a strong sub-linearity at lower temperature. 
A similar field-induced non-linearity of \Kxy~is observed at all dopings.
All lines are a guide to the eye.
}
\label{fig4}
\end{figure}

The data in Figs.~\ref{fig2} and \ref{fig3} were all taken at $H$ = 15~T.
It is worth noting that although \Kxy~is linear in $H$ at high temperature,
it develops a sublinear dependence at lower temperature, in tandem with 
the growth of a field dependence in \Kxx, as shown in Fig.~\ref{fig4} for $x=0.04$.

%
%

%
%

\begin{figure}[t]
\centering
\includegraphics[width = 0.8\linewidth]{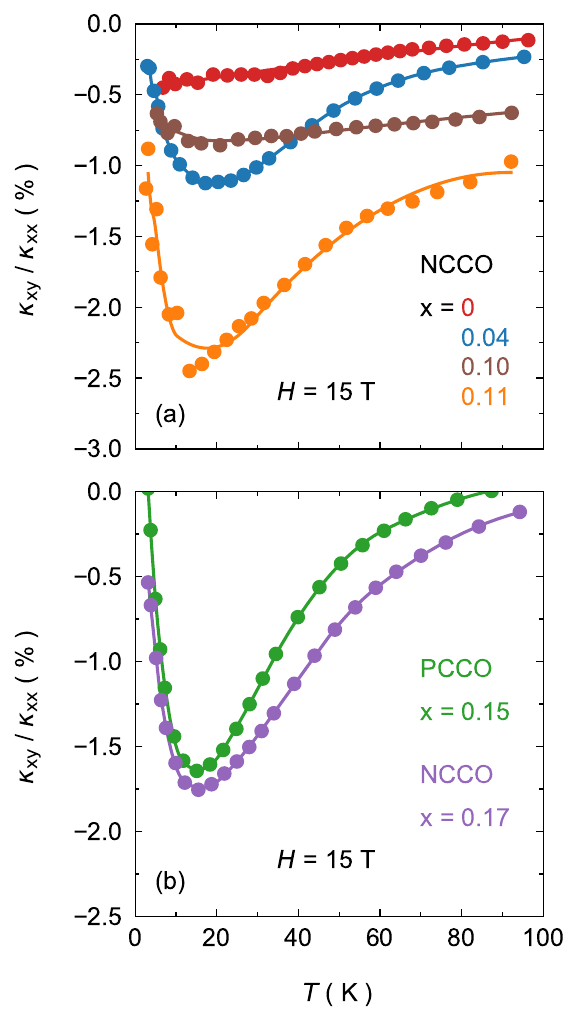}

\caption{
Ratio of \Kxy~over \Kxx,
as a function of temperature, measured at $H = 15$~T.
(a) For samples in the AF regime (data from Fig.~\ref{fig2}).
(b) For samples in the {metallic} regime (data from Fig.~\ref{fig3}).
The ratio $|$\Kxy/\Kxx$|$ -- 
a dimensionless quantity we call the degree of chirality -- 
is seen to peak at $T \simeq 20$~K, roughly
the same temperature at which the phonon conductivity \Kxx~peaks.
Its maximal value in the electron-doped samples ($x >0$) is approximately
1-3~\% (see Table~I for precise values), a factor 4 larger than in hole-doped cuprates (Table~I).
All lines are a guide to the eye.
}
\label{fig5}
\end{figure}

While for $x < 0.13$ we expect to have a negligible electronic contribution to \Kxx~and \Kxy~(because the samples are electrically insulating),
this may not be the case at the highest dopings, where the samples are reasonably good metals.
In Fig.~\ref{fig3}, we plot a rough estimate of the expected contribution of electrons (dashed lines),
calculated by assuming that the Wiedemann-Franz law holds at all temperatures,
namely
$\kappa^{\rm e}_{\rm {xx}} =   L_0   T\sigma_{\rm{xx}}$, with $\sigma_{\rm{xx}} = \rho_{\rm{xx}} / \left( \rho^2_{\rm{xx}} + \rho^2_{\rm{xy}} \right)$,
and
$\kappa^{\rm e}_{\rm {xy}} =   L_0   T\sigma_{\rm{xy}}$, with $\sigma_{\rm{xy}} = \rho_{\rm{xy}} / \left( \rho^2_{\rm{xx}} + \rho^2_{\rm{xy}} \right)$.
For these estimates, we used published data on PCCO at $x = 0.17$ --
$\rho_{\rm {xx}} (T) $ data in ref.~\onlinecite{Tafti2014}
and $R_{\rm {H}} (T)$ data in ref.~\onlinecite{Charpentier2010}, 
with $\rho_{\rm {xy}} (T) \equiv R_H (T) \times H$. 
This gives an upper bound on the electronic \Kxx~and \Kxy.
We see that these electronic contributions are small compared to the measured \Kxx~and \Kxy, especially at low temperature,
being less than 10~\% at $T <40$~K.

It is instructive to look at the ratio \Kxy/\Kxx, 
plotted as a function of temperature in Fig.~\ref{fig5}.
We see that the quantity $|$\Kxy/\Kxx$|$, which we call the degree of chirality,
peaks at $T \simeq 20$~K, the temperature where the phonon conductivity \Kxx~also peaks.
At its peak, the degree of chirality in electron-doped cuprates reaches a maximal value
$|$\Kxy/\Kxx$| = 0.9 - 2.7 \%$, at $H$= 15~T (Table~I).
This degree of chirality is some 4 times larger than that measured in the hole-doped cuprates, namely
$|$\Kxy/\Kxx$| = 0.2 - 0.6 \%$ at $H$= 15~T (Table~I).
This significant difference in the degree of chirality might provide new insight into the underlying mechanism
of phonon chirality in all cuprates, as we discuss below.


\setlength{\tabcolsep}{11pt}
\begin{table}[!]
  \caption{
  Thermal Hall conductivity of cuprates.
  The magnitude of \Kxx~and \Kxy~at $T$ = 20~K and $H$ = 15~T is listed, 
  as well as the degree of chirality, 
  given by the ratio $|$\Kxy/\Kxx$|$.
The first group of materials (top section) consists of cuprate Mott insulators.
The second group (middle section) consists of hole-doped cuprates, for doping values $p$
as indicated (second column). 
At high doping ($p > 0.1$), the samples are metallic and so we quote here the thermal transport coefficients 
for a heat current normal to the CuO$_2$ planes ($J \parallel c$), 
which contain only the phonon contribution to heat transport.
The last group is electron-doped cuprates (bottom section), for doping values $x$
as indicated.
  }
  \centering
  \label{Table:Table1}
  \setlength{\tabcolsep}{3.5pt}
  \begin{tabular*}{\linewidth}[t]{l r c c c}
    \hline
    \hline
    \noalign{\vskip 0.1cm}
    \multirow{2}{*}{Material} & \multirow{2}{*}{Doping}
     &\Kxy & \Kxx &  $\lvert \kappa_{xy} /\kappa_{xx} \rvert$ \\
    \noalign{\vskip 0.05cm}
     & &(mW/Km) & (W/Km) & (\%) \\
    \noalign{\vskip 0.1cm}
    \hline
    \noalign{\vskip 0.1cm}
    Nd$_2$CuO$_4$~{\cite{Boulanger2020}} & 0.00 & -- 200 &  56 & 0.4 \\
    Sr$_2$CuO$_2$Cl$_2$~\cite{Boulanger2020} & 0.00 & -- 21 & 7 & 0.3 \\
    La$_2$CuO$_4$~{\cite{Grissonnanche2019}} & 0.00 & -- 38 & 12 &  0.3 \\
    La$_2$CuO$_4$\footnote{\label{footnote1}($J \parallel c$)}{\cite{Grissonnanche2020}} & 0.00 & -- 30 & 16 &  0.2 \\
     \noalign{\vskip 0.1cm}
    \hline
    \noalign{\vskip 0.1cm}
    LSCO~\cite{Grissonnanche2019} & $p = 0.06$ & -- 33 & 6 & 0.6 \\
    Eu-LSCO~\cite{Grissonnanche2019} & 0.08 & -- 11 & 5 & 0.2 \\
    Nd-LSCO\footnoteref{footnote1}~{\cite{Grissonnanche2020}} & 0.21 & -- 14  & 2.9 & 0.5 \\
    Nd-LSCO\footnoteref{footnote1}~{\cite{Grissonnanche2020}} & 0.24 & 0 & 1.2 & 0 \\
    Eu-LSCO\footnoteref{footnote1}~{\cite{Grissonnanche2020}} & 0.24 & 0 & 1.2 & 0\\
    \noalign{\vskip 0.1cm}
    \hline
    \noalign{\vskip 0.1cm}
    NCCO & $x = 0.04$ & -- 314   &  28  & 1.1   \\
    NCCO\footnote{\label{footnote2}(as grown)}  & $0.04$ & -- 1400   & 51  & 2.7   \\
    NCCO\footnoteref{footnote1}    & 0.04 & -- 227 &  14 & 1.6\\
    NCCO & 0.10 & -- 194   &  22  & 0.9   \\
    NCCO\footnoteref{footnote2}    & 0.10 & -- 360 & 25 & 1.4\\
    NCCO & 0.11 & -- 898   &  39  & 2.3   \\
    PCCO & 0.15 & -- 676   &  43  & 1.6   \\
    NCCO & 0.17 & -- 568   &  34  & 1.7   \\
 \noalign{\vskip 0.1cm}
     \hline
     \hline
  \end{tabular*}
\end{table}


\begin{figure}[t]
\centering
\includegraphics[width = \linewidth]{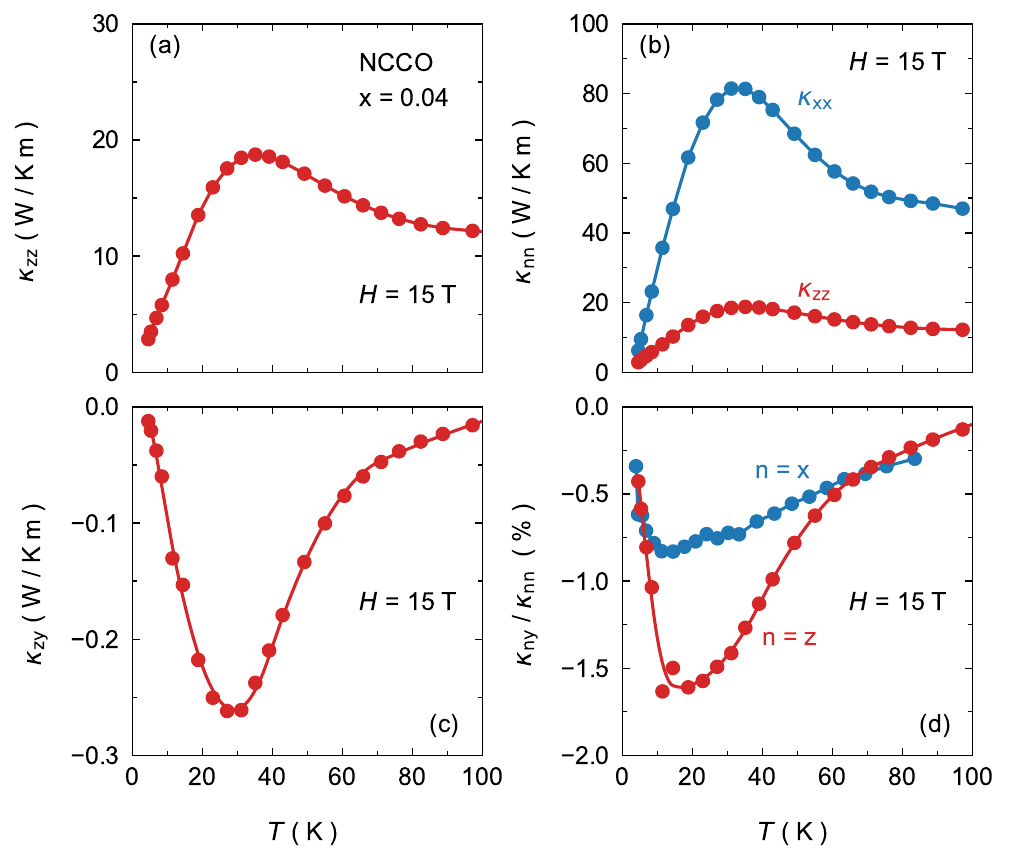}
\caption{ 
Heat transport properties of NCCO at $x = 0.04$, 
for ($J \parallel c$) ($n=z$, red)
and 
for $(J \parallel a)$ ($n=x$, blue),
at $H = 15$~T (where $H \perp J$).
(a)
$c$-axis thermal conductivity,
plotted as \Kzz~vs $T$. 
%
(b) 
Comparison of the longitudinal thermal conductivities
for a heat current in-plane (\Kxx, blue) and out-of-plane (\Kzz, red). 
%
%
The two data sets come from two samples cut from the same crystal. 
(c) 
Thermal Hall conductivity for a heat current along the $c$ axis and a field $H = 15$~T applied parallel to the CuO$_2$ planes,
plotted as \Kzy~vs $T$. 
(d) Ratio of \Kzy~over \Kzz~($n = z$, red; data from panels a and c) compared to the ratio 
of \Kxy~over \Kxx~($n = x$, blue; data taken on a separate sample cut from the same crystal).
%
%
}
\label{fig6}
\end{figure}

\begin{figure}[t]
\centering
\includegraphics[width = \linewidth]{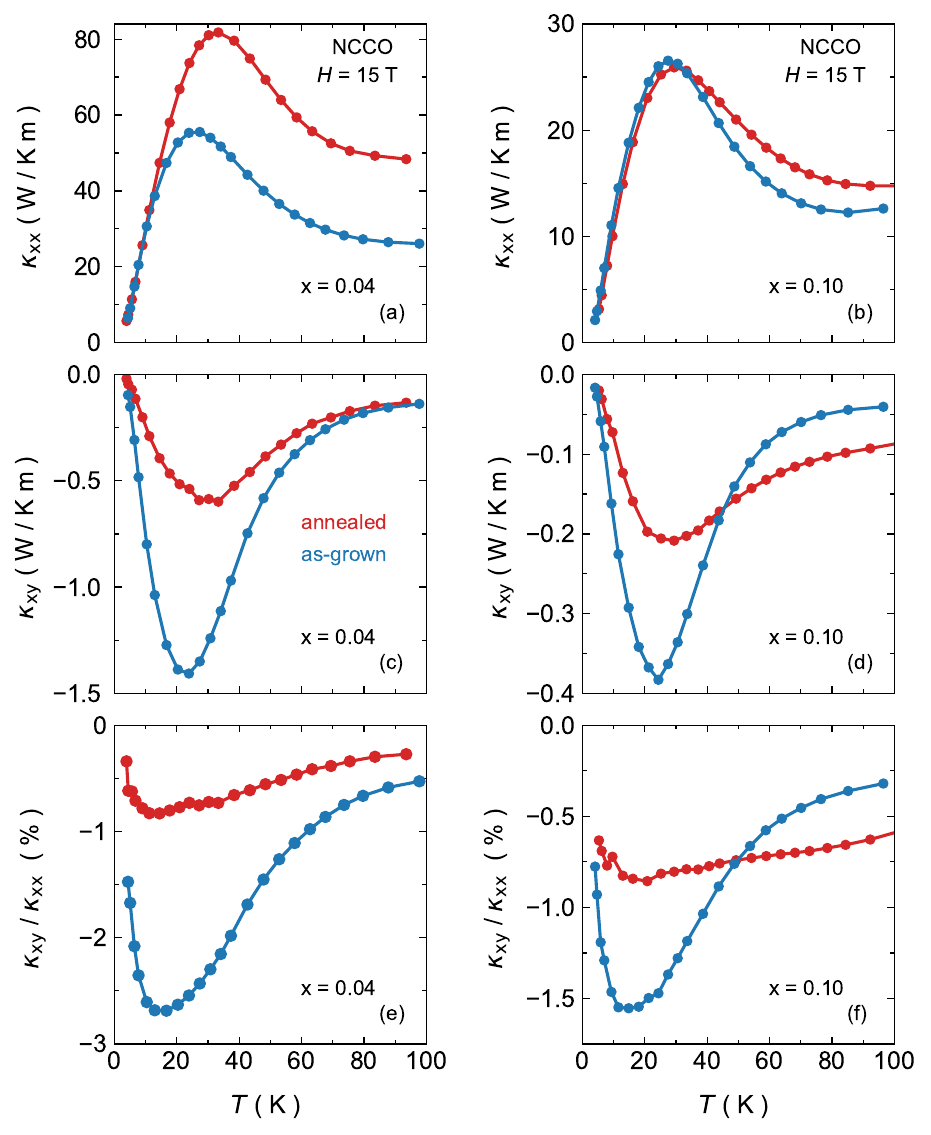}
\caption{Same as for Fig.~\ref{fig6}, but for $x$ = 0.17.
}
\label{fig7}
\end{figure}

In order to confirm that phonons are indeed the heat carriers responsible for the negative thermal Hall effect
in NCCO and PCCO,
we investigated the 
thermal Hall response for a heat current sent along the $c$ axis, normal to the CuO$_2$ planes.
In such a direction, the only excitations in the material that have non-negligible mobility are the phonons,
as argued in Ref.~\onlinecite{Grissonnanche2020}.
In Fig.~\ref{fig6}a,
we plot \Kzz~vs $T$ for NCCO at $x = 0.04$ ($J \parallel c$).
We find that \Kzz~is significantly smaller than \Kxx~($J \parallel a$),
by a factor 4 or so at $T =$~30-40~K,
where the two conductivities were measured on two separate samples cut from the same crystal (Fig.~\ref{fig6}b).
In Fig.~\ref{fig6}c,
the thermal Hall conductivity \Kzy~is displayed ($J \parallel c$, $H \parallel a$).
The degree of chirality for $J \parallel c$, $|$\Kzy/\Kzz$|$,
is seen to be even larger than that for $J \parallel a$, {\it i.e.} $|$\Kxy/\Kxx$|$ (Fig.~\ref{fig6}d).
This confirms that phonons are responsible for the negative thermal Hall effect in electron-doped cuprates.
Similar findings were obtained at $x = 0.17$ (see Fig.~\ref{fig7}).


Finally, 
we have investigated the effect of annealing on the NCCO samples,
by comparing data from as-grown samples $vs$ annealed samples at the same doping,
for two dopings: $x = 0.04$ and $x = 0.10$.
In its ideal structure, there is no oxygen atom in the 
apical position of the crystal lattice of NCCO.
However, in as-grown crystals there is an oxygen atom in approximately 10~\% of apical positions~\cite{Radaelli1994}. The reduction annealing treatment 
removes some of those excess oxygens in apical positions, especially when 
$x \leq$ 0.10~\cite{Kang2007,Richard2004}. 
Annealing crystals also reduces built-in tension and disorder.

\begin{figure}[t]
\centering
\includegraphics[width = \linewidth]{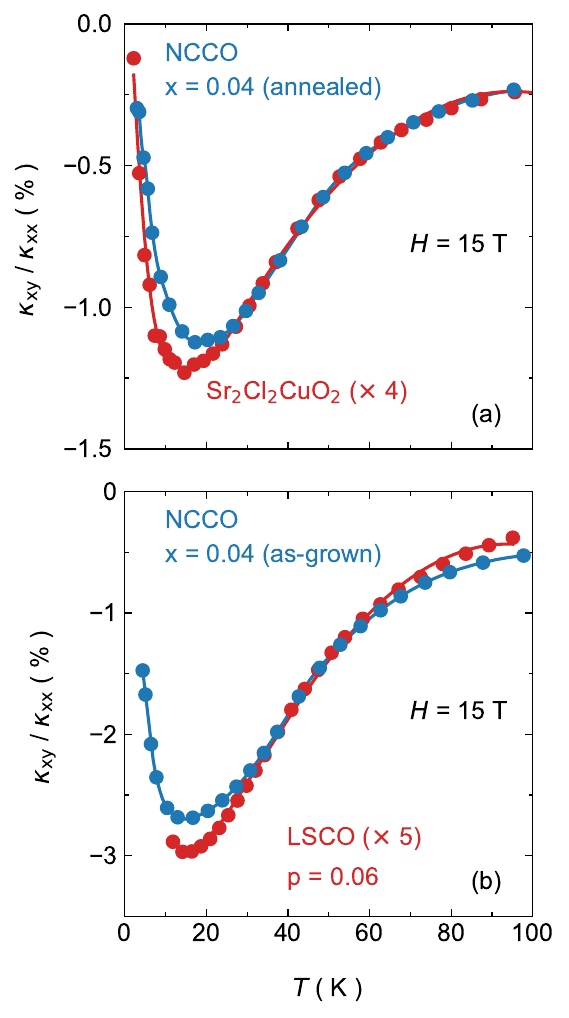}
\caption{
Effect of annealing on the thermal transport of NCCO
at 
$x = 0.04$ (left panels) 
and 
$x = 0.10$ (right panels). 
Data taken on an as-grown (not annealed) sample (blue curves) are compared to data taken
on an annealed sample (red curves):
(a,b) longitudinal thermal conductivity \Kxx;
(c,d) thermal Hall conductivity \Kxy;
(e,f) ratio \Kxy/\Kxx.
}
\label{fig8}
\end{figure}

The thermal transport data for this comparative study are displayed in Fig.~\ref{fig8}. 
Let us first focus on the data for the sample with $x=0.04$ (left panels).
In Fig.~\ref{fig8}a,
we see that the removal of excess oxygens by annealing has little effect on \Kxx~at the lowest temperatures ($T < 10$~K),
a regime where the sample boundaries dominate the scattering of phonons,
but it leads to a significant increase in the phonon mean free path at higher temperatures,
seen as an enhanced conductivity \Kxx.
%

Surprisingly, 
this enhanced conductivity is accompanied by a {\it decrease} of $|$\Kxy$|$~(Fig.~\ref{fig8}c), 
a behavior opposite to the usual trend that the magnitude of \Kxy~tends to grow in tandem with \Kxx~(Table~I).
As a result, the degree of chirality -- $|$\Kxy/\Kxx$|$ -- is seen to be significantly larger in the as-grown sample~(Fig.~\ref{fig8}e).
We discuss the possible implications of this observation in the next section.

Corresponding data for the sample with $x = 0.10$ (Fig.~\ref{fig8}, right panels),
do not lead to as clearcut a conclusion, because the effect of annealing is different at higher doping -- indeed, apical oxygens are no longer affected, but in-plane oxygens are removed~\cite{Riou2004}. Consistent with the resulting in-plane disorder, a prior study found reduction annealing of NCCO at $x=0.22$ to decrease \Kxx~\cite{Cohn1992}.
Nonetheless, at its peak $vs$ temperature, $|$\Kxy$|$~is again larger in the as-grown sample (Fig.~\ref{fig8}d),
as is $|$\Kxy/\Kxx$|$ (Fig.~\ref{fig8}f).

\section{DISCUSSION}

We observe 
a large negative thermal Hall conductivity \Kxy~in the electron-doped cuprates NCCO and PCCO,
for all dopings from the insulator at $x=0$ to the metal at $x = 0.17$.
After summarizing the arguments for why phonons are the heat carriers responsible for this \Kxy~signal,
we discuss the implications of our findings regarding the
underlying mechanism that makes phonons chiral in cuprates -- 
by which we simply mean that they produce a thermal Hall effect in the presence of a magnetic field perpendicular to the heat current.

\subsection{
Phonon Hall effect}

The similarity of \Kxy~data at the various dopings {strongly suggests}
a single unified scenario -- 
the same heat carriers and the same {microscopic mechanism at all dopings}.
The negative \Kxy~signal we detect is certainly not due to electrons at low doping,
for our sample with $x = 0.04$ is electrically insulating. 
Nor is it due to magnons at high doping, for our samples at $x = 0.15$ and $x = 0.17$
are outside the phase of long-range AF order.
So we can rule out electrons and magnons as the heat carriers responsible for the thermal Hall effect in 
electron-doped cuprates across the full doping range.This leaves phonons as the only obvious candidate.
The fact that the curve of \Kxy~vs $T$ peaks at the same temperature as the phonon-dominated \Kxx~vs $T$ certainly 
suggests that phonons are responsible for \Kxy.
(Note that in hole-doped cuprates (LSCO, EU-LSCO), \Kxy~vs $T$ also peaks at $T \sim 20$~K, which is typically the temperature at which the phonon thermal conductivity \Kxx~peaks~\cite{Grissonnanche2019}.)

What essentially proves that it is phonons that carry the thermal Hall effect is our study of heat transport along the $c$ axis,
a direction in which only phonons can carry heat with significant mobility,
where we also observe a large 
degree of chirality (Figs.~\ref{fig6}d and \ref{fig7}d).

\subsection{
Scattering processes}

It is likely that the phonon thermal Hall effect in cuprates 
is due to the skew scattering of phonons off some impurities or defects.
The original observation of a phonon Hall effect, in the insulator Tb$_3$Ga$_5$O$_{12}$~\cite{Strohm2005},
was attributed to skew scattering off superstoichiometric Tb$^{3+}$ ions~\cite{Mori2014}.
In SrTiO$_3$,
the large \Kxy~signal was linked to the presence of structural domains, resulting from
a structural transition at $T \simeq 105$~K~\cite{Li2020,Chen2020},
thereby accounting for the very weak \Kxy~signal seen in the closely related material KTaO$_3$,
which does not undergo any such transition.
Let us therefore explore what scattering processes might be relevant for electron-doped cuprates.

The first point to stress is that there are no structural domains in NCCO or PCCO,
as these materials retain their tetragonal structure at all temperatures.
So scattering of phonons off structural domain boundaries is ruled out
(including those associated with the AF order, as there is no long-range order in our samples with $x = 0.15$ and 0.17.)

The next observation is the striking similarity between \Kxy~data on NCCO and PCCO~(Figs.~\ref{fig3}b and \ref{fig5}b).
This immediately tells us that the nature of the rare-earth ion, whether Nd$^{3+}$ or Pr$^{3+}$, 
is irrelevant for the mechanism of phonon chirality in cuprates.

The larger degree of chirality in NCCO with $x > 0$ compared to Nd$_2$CuO$_4$~($x = 0$)
suggests that adding Ce ions into the structure favours phonon chirality.
Could it be that the scattering of phonons by Ce impurities enhances the \Kxy~signal in cuprates?
This possibility -- that Ce atoms are effective skew scatterers of phonons in cuprates -- 
should be tested deliberately, in future work. Note that \Kxx~does decrease with increasing Ce content (Fig.~\ref{fig2}), as observed in a prior study~\cite{Cohn1992}. (The fact that \Kxx~is larger at $x = 0.11$ than at $x = 0.04$ is presumably due to a lower level of other disorder in the former.)

Our study of the effect of annealing revealed that excess oxygen atoms on apical locations in the lattice
scatter phonons (decrease \Kxx) and enhance the degree of chirality (increase $|$\Kxy/\Kxx$|$)-- at least at $x = 0.04$ (Fig.~\ref{fig8}).
This points to another potential mechanism for skew scattering of phonons that would be worth investigating theoretically.
It has been proposed that charged defects in ionic crystals can produce a skew scattering of phonons 
and thus yield a thermal Hall effect in such insulators~\cite{flebus2021}.
However, this particular mechanism -- associated with a local charge -- seems unlikely for NCCO 
since the degree of chirality remains large even in the metallic regime, 
where conduction electrons would surely screen any local charge. 

Of course, there are other impurities and defects in NCCO and PCCO beyond Ce ions and apical oxygens.
The question is what types of defects -- and in what environment -- will produce the necessary skew scattering.
This is being explored theoretically ({\it e.g.}~\cite{Guo2021,sun2021,guo2022}) and empirically ({\it e.g.}~\cite{Chen2021}),
as a growing number of insulators are found to exhibit a sizable phonon thermal Hall effect.
Now because defects and impurities in these various materials must be widely different,
there seems to be something rather general about the skew scattering of phonons in a magnetic field that needs to be understood.

This is of course not universal, since there are insulators that have \Kxy~$=0$, like
Y$_2$Ti$_2$O$_7$~\cite{Hirschberger2015} and LiF~\cite{Kasahara2018}, for example.

We propose that the mechanism of phonon chirality might involve the combination of 
local defects and local spins, or short-range magnetic correlations.
The defects would locally distort the spin environment and generate a local spin chirality
that would result in skew scattering of phonons when a field is applied.
This would imply that non-magnetic insulators would not have a phonon Hall effect.
(An immediate counter example is SrTiO$_3$, and so for this particular quantum paraelectric material,
another mechanism would need to be invoked -- presumably involving high electric polarizability 
and structural domain boundaries.)

\begin{figure}[t]
\centering
\includegraphics[width = 0.8\linewidth]{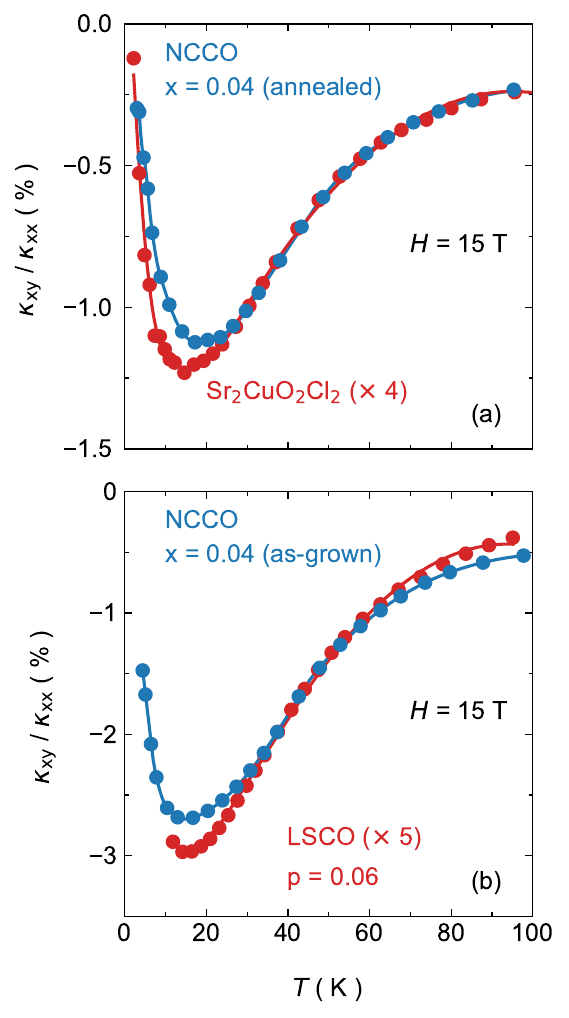}

\caption{
(a) Comparison of the ratio \Kxy /\Kxx~for NCCO $x=0.04$ (annealed, blue; this work) 
and the cuprate Mott insulator Sr$_2$CuO$_2$Cl$_2$ (red;~\cite{Boulanger2020}), 
in a magnetic field $H = 15$~T. 
The data for Sr$_2$CuO$_2$Cl$_2$ are multiplied by a factor 4. 
(b) Similar comparison between NCCO $x=0.04$ (as-grown, blue; this work)
and the hole-doped cuprate LSCO with doping $p = 0.06$ (red;~\cite{Grissonnanche2019}).
The data for LSCO are multiplied by a factor 5.
Lines are a guide to the eye.
}
\label{fig9}
\end{figure}

\subsection{
Electron-doped vs hole-doped cuprates}

It is instructive to compare electron-doped cuprates with hole-doped cuprates.
Both exhibit a negative phonon thermal Hall effect,
which is independent of whether the sample is inside or outside the regime of long-range AF order.
The temperature dependence of \Kxy~is very similar in the two families of cuprates,
as illustrated in Fig.~\ref{fig9}.

There are two important differences.
First, a quantitative difference: 
the magnitude of $|$\Kxy/\Kxx$|$ is significantly larger in electron-doped cuprates,
by a factor 4 or so (Table~I).
We see three possible ingredients that may be relevant.
Electron-doped cuprates contain Ce atoms, which may act as skew scatterers.
Also, they do not have apical oxygens in their pristine structure,
so impurity atoms in those locations -- present even in annealed samples -- act as scattering centers,
which, as argued above, appear to increase the degree of phonon chirality.
Finally, antiferromagnetism is more pronounced in electron-doped cuprates,
in the sense that the phase of long-range AF order extends to much higher doping -- 
up to $x \simeq 0.13$ in NCCO vs $p \simeq 0.02$ in LSCO -- 
and short-range AF correlations
in NCCO persist beyond that limit, up to $x =0.17$~\cite{motoyama2007}.
We suggest that the role of such AF correlations in causing phonon chirality is a promising avenue 
of investigation.
In summary, a quantitative comparison between electron-doped and hole-doped cuprates
points to three possible factors to account for the stronger chirality in the former:
Ce impurities, excess apical oxygens and stronger AF correlations. 

Note that another difference between electron-doped and hole-doped cuprates is the field dependence of \Kxx~and \Kxy, shown for NCCO in Fig.~\ref{fig4}. In NCCO at $x = 0.04$, and also at other dopings, the field dependence of \Kxx~is much stronger than in hole-doped cuprates~\cite{Grissonnanche2019}. The field suppresses \Kxx, suggesting a field induced enhancement of phonon scattering. This contrasts with the weak dependence seen in hole-doped cuprates and the slight increase of \Kxx~vs $H$ seen in Sr$_2$CuO$_2$Cl$_2$~\cite{Boulanger2020}. As for \Kxy, all cuprates show the same qualitative behaviour: \Kxy~is linear in $H$ at high temperature and sublinear at low temperature. However, this sublinearity is particularly strong in Nd$_2$CuO$_4$~\cite{Boulanger2020} and electron-doped cuprates. These effects may guide theoretical developments.

The second important difference is qualitative.
In NCCO, a sizable negative \Kxy~signal persists up to the highest doping ($x = 0.17$), but
this is not the case in the hole-doped materials Nd-LSCO and Eu-LSCO:
at a doping $p = 0.24$, just above the pseudogap critical point at $p^{\star} = 0.23$~\cite{Daou2009,Collignon2017},
there is no detectable phonon thermal Hall effect~\cite{Grissonnanche2019}.
In other words, phonons cease to be chiral outside the pseudogap phase --
a negative phonon \Kxy~signal is only present in these materials when $p < p^{\star}$~\cite{Grissonnanche2019}.
We infer that some intrinsic property of the pseudogap phase,
also present in the Mott insulating state (at $p=0$),
is needed to confer chirality to phonons.
It remains to be seen whether the phonon thermal Hall effect in electron-doped cuprates would also disappear at dopings above $x = 0.17$, were single crystals available for measurements at such high dopings.

Considering all aspects of this discussion, 
we are led to propose that the mechanism of phonon chirality in cuprates,
responsible for the phonon thermal Hall effect in both hole-doped and electron-doped materials,
relies on the combination of scattering by defects and short-range AF correlations.
In other words,
the scattering of phonons by defects becomes skew scattering (in a field)
when there are local spins at or near the defect location.
This would imply that short-range AF correlations are present in the pseudogap phase of hole-doped cuprates 
but absent (or weak) outside, as indeed found in recent experimental studies of LSCO~\cite{Frachet2020}. This picture is also consistent with numerical solutions of the Hubbard model that find a pseudogap phase characterized by short-range AF correlations up to a critical doping $p^{\star}$~\cite{Sordi2010}.


\section{SUMMARY}

We have measured the thermal conductivity \Kxx~and the thermal Hall conductivity \Kxy~of the electron-doped cuprates NCCO and PCCO
across their phase diagram. 
We observe a large negative thermal Hall conductivity over the whole doping range, 
from $x = 0$ in the Mott insulating state up to at least $x = 0.17$ in the metallic state. 
We show that this negative signal is carried by phonons.

We find that the degree of chirality, measured by the ratio $|$\Kxy/\Kxx$|$,
is enhanced by addition of Ce to Nd$_2$CuO$_4$
and by having excess oxygen atoms at apical locations of the NCCO lattice.
Both of these defects therefore appear to be effective skew scatterers.

In comparison to hole-doped cuprates, 
electron-doped cuprates have a significantly larger ratio $|$\Kxy/\Kxx$|$,
perhaps because neither of the two skew scattering processes just mentioned
are present in the former.
The fact that a sizable phonon \Kxy~signal persists up to the highest measured doping
in electron-doped NCCO ($x = 0.17$) but vanishes above the critical doping $p^{\star}$ in hole-doped Nd-LSCO
points to an intrinsic property of the pseudogap phase in the latter, only present below $p^{\star}$.
We suggest that this property could be short-range AF correlations.

\section{ACKNOWLEDGMENTS}

We thank L. Balents, M. Dion, P. Fournier, C. Gauvin-Ndiaye, S.A. Kivelson, S. Sachdev and A.-M.S. Tremblay for fruitful discussions.
We thank S.~Fortier for his assistance with the experiments.
L.T. acknowledges support from the Canadian Institute for Advanced Research and funding from
the Institut Quantique,
the Natural Sciences and Engineering Research Council of Canada (Grant No, PIN:123817),
the Fonds de Recherche du Qu\'{e}bec - Nature et Technologies,
the Canada Foundation for Innovation,
and a Canada Research Chair.
This research was undertaken thanks in part to funding from the Canada First Research Excellence Fund.
Z-X.S. acknowledges the support of the U.S. Department of Energy, Office of Science, Office of Basic Energy Sciences, Division of Material Sciences and Engineering, under contract DE-AC02-76SF00515.
R.L.G acknowledges the support of the National Science Foundation under award DMR 2002658.

\vfill

\bibliography{reference}

\end{document}